# Reduced Exchange Interactions in Magnetic Tunnel Junction Free Layers with Insertion Layers

Jamileh Beik Mohammadi,[1] Bartek Kardasz,[2] Georg Wolf,[2] Yizhang Chen,[1] Mustafa Pinarbasi,[2] and Andrew D. Kent[1]

[1]Center for Quantum Phenomena, Department of Physics, New York University, New York, NY 10003, USA

[2]Spin Memory, Inc., Fremont, California, CA 94538, USA

## Abstract

Perpendicularly magnetized CoFeB layers with ultra-thin non-magnetic insertion layers are very widely used as the electrodes in magnetic tunnel junctions for spin transfer magnetic random access memory devices. Exchange interactions play a critical role in determining the thermal stability of magnetic states in such devices and their spin torque switching efficiency. Here the exchange constant of free layers incorporated in full magnetic tunnel junction layer stacks, specifically CoFeB free layers with W insertion layers is determined by magnetization measurements in a broad temperature range. A significant finding is that the exchange constant decreases significantly and abruptly with W insertion layer thickness. The perpendicular magnetic anisotropy shows the opposite trend; it initially increases with W insertion layer thickness and shows a broad maximum for approximately one monolayer (0.3 nm) of W. These results highlight the interdependencies of magnetic characteristics required to optimize the performance of magnetic tunnel junction devices.

**Main text**

Magnetic tunnel junctions with perpendicularly magnetized layers (pMTJs) are of great interest for high-density memories such as in magnetic random access memory (MRAM) devices.[1] An optimized free layer is an essential part of the pMTJ and is designed to attain high thermal stability of its magnetic states, while permitting fast and low voltage spin transfer torque switching. The free layer is typically a thin body-centered cubic (bcc) CoFeB layer with an interface to MgO, which creates a large tunneling magnetoresistance.[2, 3, 4] The required thermal stability for long term data storage is achieved through interface perpendicular magnetic anisotropy with adjacent MgO layers on both interfaces.[3, 4] Ultrathin W and Ta insertion layers can further increase the perpendicularly magnetic anisotropy and thus lead to longer thermal stability.[3, 5]

The switching dynamics and the thermal stability of the magnetic states—the energy barrier to thermally activated magnetization reversal ($E_b$)—depend on the magnetic anisotropy and the exchange constant of the free layer.[6-8] The latter is important because magnetization reversal can occur through non-uniform magnetic states, such as domain nucleation and domain wall propagation. The exchange constant, along with effective perpendicular anisotropy $K_{eff}$, set the length scale below which coherent magnetization dynamics are expected (a critical diameter, $d_c$, for a free layer element in the shape of a very thin disk), as well as the energy barrier for thermally activated magnetization reversal for free layer elements greater than this length scale.[7, 8] Several studies have characterized the perpendicular anisotropy of free layer materials,[9] but the exchange constant of pMTJ free layers has not been systematically investigated. Earlier studies suggest that the exchange constant is smaller for thin films compared to the bulk materials.[11, 12] However, the exchange stiffness of CoFeB free layers—which are widely used in MRAM—and the effects of an ultrathin insertion layer on the exchange constant have not been determined.

The exchange constant can be determined using dynamic methods such as Brillion light scattering (BLS)[13] and spin-transfer ferromagnetic resonance (ST-FMR).[11,14] However, these methods have significant limitations when applied to the study of magnetic tunnel junction materials and devices. For example, BLS cannot typically be used to study ultrathin films. Further, estimation of the exchange constant from ST-FMR measurement requires identifying the spin-wave modes that are excited, which can be complicated due to (1) the selection rules involved,[15] (2) the selection of proper boundary conditions, (3) the presence of defects, and (4) fabrication related changes in material properties at the edges of the device. Moreover, these methods do not generally allow characterization of the exchange constant of the other magnetic layers in a multilayer stack. Such limitations are lifted through the use of precise quasi-static magnetization measurements, as a function of temperature, that probe the temperature dependence of the magnon population. This method allows one to determine the exchange constant of individual magnetic layers in MRAM layer stacks when the magnetic layers have well-separated switching fields.

We have determined the exchange constant in dual MgO CoFeB free layers with W insertion layer by magnetization measurements in a broad temperature range. A significant finding is that the exchange constant decreases significantly and abruptly with W insertion layer thickness, while the perpendicular magnetic anisotropy shows the opposite trend; it initially increases with W insertion layer thickness and shows a broad maximum at about one monolayer (0.3 nm) of W. These results highlight interdependencies of the magnetic properties that have to be considered when optimizing the layer configurations in pMTJs for advanced applications.

We studied full pMTJ layer stacks comprised of SAF1/Ru/SAF2/MgO/CoFeB/W($t_W$)/CoFeB /MgO, illustrated schematically in **Figure 1**a. SAF1 and SAF2 form the synthetic antiferromagnet, which is used to stabilize the reference layer in the pMTJ and reduce the internal stray fields acting on the free layer magnetization in a device. The free layer is formed by CoFeB/W($t_W$)/CoFeB and is sandwiched between the main MgO tunnel barrier and a MgO cap layer. The two CoFeB layers (specifically, $Co_{18}Fe_{54}B_{28}$ layers) with a total thickness of 2.3 nm are coupled ferromagnetically through the W layer. The main focus of this experiment is on the effects of the ultrathin W insertion layer on the free layer's magnetic properties. The thickness $t_W$ is varied between 0.1 and 0.7 nm while the other layers are not changed. In addition, we have studied a single CoFeB layer with MgO interfaces, a MgO/CoFeB/MgO layer stack, to compare its magnetic properties to those of the dual MgO CoFeB composite free layers.

We performed vibrating sample magnetometry (VSM) in a field-perpendicular geometry in a temperature range of 5 to 395 K. Wafers were diced into squares so that samples of precisely the same size and shape could be studied and compared. This minimizes the error introduced when determining the sample volume and thus the uncertainty in the saturation magnetization. Two samples were glued face to face to improve the signal to noise ratio (to double the magnetic moment). Figure 1b presents the major and minor hysteresis loop data for a pMTJ stack with a 0.2 nm W ($t_W$=0.2 nm) insertion layer. Different colors represent different temperatures in the range of 5 to 380 K.

A challenge when studying a full pMTJ stack is to determine the full saturation moment of each layer. At each temperature, the saturation moment of individual layers is determined to be the amplitude of the corresponding steps in the major hysteresis loops. The steps are shown and labeled in Figure 1b. The free layer (FL) has the smallest switching field, less than about 0.05 T. SAF1 and SAF2 are antiferromagnetically coupled and have larger perpendicular magnetic anisotropy leading to switching fields close to 0.5 T. In addition to using the step in the major hysteresis loops, the free layer's magnetic moment is also determined by subtracting the saturation moments of the SAF layers from the total saturation moment. In a third method, the magnetic moment of the free layer is extracted from minor hysteresis loops in the same range of temperatures (see the inset of Figure 1b). The magnetization is obtained by dividing the saturation moment by the layer volume.

The temperature dependence of the magnetization can be determined within a Heisenberg model with Hamiltonian $H = -J\sum_{<ij>} \widehat{S}_i \cdot \widehat{S}_j$, where $J$ is the exchange energy and $\widehat{S}_i, \widehat{S}_j$ are spin operators associated with sites $i$ and $j$. This model predicts the well-known Bloch's $T^{3/2}$ law.[16] Well below the Curie temperature, the magnetization decreases as:

$$m(T) = \frac{M_s(T)}{M_s(0)} = 1 - \frac{\eta}{2S}\left(\frac{k_B T}{2SJ}\right)^{3/2}, \qquad (1)$$

where $M_s$(T) is the magnetization at temperature $T$, $S$ is the spin at each lattice site (in units of Planck's constant, $\hbar$), $k_B$ is Boltzmann's constant, and $\eta$ is a dimensionless constant equal to 0.58 for a bulk ferromagnet; $\eta$ is modified in thin films with magnetic anisotropy (as discussed in the supplementary section).

We thus analyze our data by plotting the normalized magnetization *m* versus $T^{3/2}$. This is shown in **Figure 2** for a MgO/2.3 nm CoFeB/MgO single magnetic layer and pMTJ free layers with 0.1-0.3 nm thick W insertion layers. The data follows a straight line when the magnetization is plotted versus $T^{3/2}$, consistent with Bloch's law.

In a continuum model, the zero-temperature exchange constant *A* for a bcc lattice is related to *J* as $A = \frac{2JS^2}{a}$, where *a* is the lattice constant.[16] *S* is given in terms of the magnetization as $S = \frac{M_s a^3}{2g\mu_B}$ where $\mu_B$ is Bohr magneton and is *g*-factor. We determine *S* from the extrapolated zero temperature magnetization of the 2.3 nm CoFeB to be 0.74. Following this procedure, we obtained $A$ =8.5 pJ m$^{-1}$ for the 2.3 nm MgO/CoFeB/MgO sample. For the CoFeB layer with a 0.3 nm thick W insertion, the exchange constant is 4.0 pJ m$^{-1}$. Using this method, we determine the exchange constant of the free layer and the SAF layers as a function of W thickness. This is shown in Figure 3a and Figure 3b. We verified our measurement and analysis methods by studying NiFe thin films that were supplied by the National Institute of Standards and Technology (NIST)[18] (see supplementary materials).

Ferromagnetic resonance spectroscopy (FMR) was used to determine the effective perpendicular anisotropy of the MgO/CoFeB/MgO sample and the pMTJ free layers with 0.1-0.4 nm insertion layers. The effective perpendicular anisotropy is the difference between the anisotropy associated with spin-orbit interactions and demagnetization effects (that favor in-plane magnetization), $K_{\text{eff}} = K_p - \frac{\mu_0 M_s^2}{2}$, where $K_p$ is the perpendicular uniaxial anisotropy and $\frac{\mu_0 M_s^2}{2}$ is the demagnetization energy. FMR measurements directly determine the effective magnetization, $M_{\text{eff}} = \frac{2K_{\text{eff}}}{\mu_0 M_s}$, where positive $M_{\text{eff}}$ implies a net perpendicular magnetic anisotropy. Measurements were performed in a frequency range of 10-30 GHz at room temperature. The effective magnetization is plotted versus $t_W$ in Figure 3c, showing that there is a net perpendicular anisotropy for 0.2, 0.3 and 0.4 nm thick W insertion layers.

Our data reveal that the exchange constant of the MgO/CoFeB/MgO samples is much lower than that for thicker CoFeB samples, i.e. 19-28 pJ m$^{-1}$.[12, 13] This indicates that the exchange constant is significantly affected by interfaces and reduced dimensions. Moreover, insertion of the W layer further reduces the exchange constant of the CoFeB layer as seen in Figure 3a. As the thickness of the W layer is increased, the exchange constant of the free layer is reduced even further, to about 4 pJ m$^{-1}$ (Figure 3a). As can be seen in Figure 3b, the exchange constants of SAF1 and SAF2 layers agree well for all the stacks that are studied here. This indicates that the properties of these layers do not change with the W insertion layer and highlights that individual layer properties can be changed without affecting the properties of other layers in the MTJ stack.

The magnetic parameters of the free layers found in this study are summarized in Table 1. We note that the magnetic moment of the free layer tends to decrease with W insertion layer thickness. The magnetization times the free layer thickness, $M_s t$, that is the magnetic moment of the free layer divided by its area, is shown in Table 1. As noted above, in determining the exchange constant, we have assumed that the spin $S$ on each lattice site is the same for each sample. The varying $M_s t$ would then indicate that the effective magnetic thicknesses decrease with W insertion or that there are lower moment Fe and Co atoms at the W interfaces (i.e. what has been termed dead layers).[5, 17] Another assumption that we consider in the supplementary section, is that the W layer reduces the magnetization of the entire CoFeB layer.

We note that Bloch's law analysis gives the zero-temperature exchange constant. However, finite temperature micromagnetic simulations should use a renormalized exchange constant $A(T) = A\, m(T)^{\gamma+1}$, with $\gamma = 1$, the mean field result.[18] This implies an even further reduction in the exchange constant at room temperature of CoFeB layers with increasing W insertion layers.

In summary, we have determined the exchange constant of dual MgO CoFeB free layers in pMTJ layer stacks. We have observed that the exchange constant of the composite free layer in a pMTJ stack is about two times lower compared to the single CoFeB layer in a MgO/CoFeB/MgO sample. This is expected to have significant consequences on MRAM performance. Most importantly, the critical diameter for coherent switching (that is $d_c = (\frac{16}{\pi})\sqrt{A/K_{\text{eff}}}$) [7] decreases significantly with increasing W insertion layers thickness. For example, for a pMTJ with 0.2 nm W insertion layer $d_c \approx 35$ nm at room temperature but this parameter decreases to 27 nm when $t_W$ is 0.3 nm. This implies that switching of a 30 nm pMTJ free layer with 0.2 nm insertion would be more coherent than the same size junctions that have 0.3 nm insertion layer. Thus, sub-volume excitation, and therefore delayed switching and lower switching efficiencies [6] are likely in large junctions with a 0.3 nm insertion layer. Moreover, the reduced exchange constant of the free layer reduces the energy barrier for thermally activated switching for junctions that are greater than $d_c$. This illustrates the importance of precise determination of the exchange constant in order to be able to optimize the materials for pMTJs for MRAM devices. It is particularly important to develop composite free layers with stronger exchange while maintaining a strong perpendicular magnetic anisotropy. In general, the method we have presented can be applied to determine the exchange constant of individual ferromagnetic layers in magnetic multilayers for which the magnetic reversal of the individual layers is well spaced in a magnetic field.

**Acknowledgements**

We thank Dr. Justin M. Shaw and Dr. Hans T. Nembach from NIST for providing the NiFe test sample. This research is supported by Spin Memory, Inc. This work was supported partially by the MRSEC Program of the National Science Foundation under Award No. DMR-1420073. A.D.K. received support from the National Science Foundation under Grant No. DMR-1610416.

## References

[1]     A. D. Kent, D. C. Worledge, *Nat. Nanotechnol.* **2015**, 10, 187.

[2]     a) S. Yakata, H. Kubota, Y. Suzuki, K. Yakushiji, A. Fukushima, S. Yuasa, K. Ando, *J. Appl. Phys.* **2009**, 105, 07D131; b) D. C. Worledge, G. Hu, D. W. Abraham, J. Z. Sun, P. L. Trouilloud, J. Nowak, S. Brown, M. C. Gaidis, E. J. O'Sullivan, R. P. Robertazzi, *Appl. Phys. Lett.* **2011**, 98, 022501.

[3]     H. Sato, M. Yamanouchi, S. Ikeda, S. Fukami, F. Matsukura, H. Ohno, *Appl. Phys. Lett.* **2012**, 101, 022414.

[4]     T. Devolder, J. Kim, J. Swerts, S. Couet, S. Rao, W. Kim, S. Mertens, G. Kar, V. Nikitin, *IEEE Trans. Magn.* **2018**, 54, 1.

[5]     J.-H. Kim, J.-B. Lee, G.-G. An, S.-M. Yang, W.-S. Chung, H.-S. Park, J.-P. Hong, *Sci. Rep.* **2015**, 5, 16903.

[6]     J. Z. Sun, R. P. Robertazzi, J. Nowak, P. L. Trouilloud, G. Hu, D. W. Abraham, M. C. Gaidis, S. L. Brown, E. J. O'Sullivan, W. J. Gallagher, D. C. Worledge, *Phys. Rev. B.* **2011**, 84, 064413.

[7]     G. D. Chaves-O'Flynn, G. Wolf, J. Z. Sun, A. D. Kent, *Phys. Rev. Appl.* **2015**, 4, 024010.

[8]     L. Thomas, G. Jan, S. Le, Y. Lee, H. Liu, J. Zhu, S. Serrano-Guisan, R. Tong, K. Pi, D. Shen, R. He, J. Haq, Z. Teng, R. Annapragada, V. Lam, Y. Wang, T. Zhong, T. Torng, P. Wang, presented at *IEEE International Electron Devices Meeting (IEDM)*, 7-9 Dec. 2015, **2015**.

[9]     J. M. Shaw, H. T. Nembach, M. Weiler, T. J. Silva, M. Schoen, J. Z. Sun, D. C. Worledge, *IEEE Magn. Lett.* **2015**, 6, 1.

[10]    C. A. F. Vaz, J. A. C. Bland, G. Lauhoff, *Rep. Prog. Phys.* **2008**, 71, 056501.

[11]    T. Devolder, *Phys. Rev. B.* **2017**, 96, 104413.

[12]     C. Bilzer, T. Devolder, J.-V. Kim, G. Counil, C. Chappert, S. Cardoso, P. P. Freitas, *J. Appl. Phys.* **2006**, 100, 053903.

[13]     A. Conca, T. Nakano, T. Meyer, Y. Ando, B. Hillebrands, *J. Appl. Phys.* **2017**, 122, 073902.

[14]     C. J. Safranski, Y.-J. Chen, I. N. Krivorotov, J. Z. Sun, *Appl. Phys. Lett.* **2016**, 109, 132408.

[15]     V. V. Naletov, G. de Loubens, G. Albuquerque, S. Borlenghi, V. Cros, G. Faini, J. Grollier, H. Hurdequint, N. Locatelli, B. Pigeau, A. N. Slavin, V. S. Tiberkevich, C. Ulysse, T. Valet, O. Klein, *Phys. Rev. B.* **2011**, 84, 224423.

[16]     A. Aharoni, *An Introduction to the Theory of Ferromagnetism*, Clarendon Press, New York, USA **2000**.

[17]     H. Honjo, S. Ikeda, H. Sato, K. Nishioka, T. Watanabe, S. Miura, T. Nasuno, Y. Noguchi, M. Yasuhira, T. Tanigawa, H. Koike, H. Inoue, M. Muraguchi, M. Niwa, H. Ohno, T. Endoh, *IEEE Trans. Magn.* **2017**, 53, 1.

[18]     H. T. Nembach, J. M. Shaw, M. Weiler, E. Jué, T. J. Silva, *Nat. Phys.* **2015**, 11, 825.

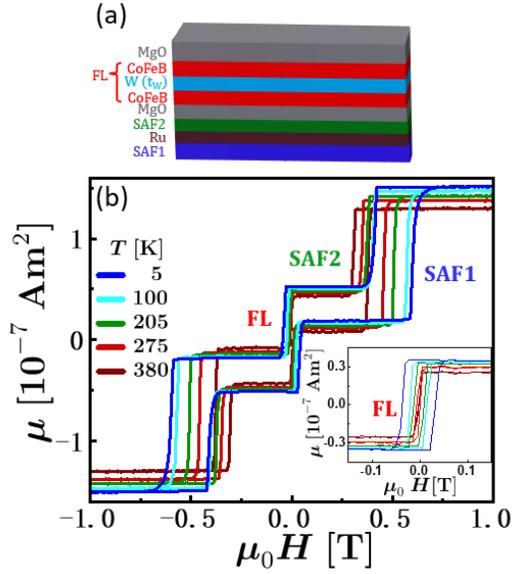

**Figure 1.** a) Perpendicular magnetic tunnel junction layer stacks studied in this paper. b) Hysteresis loops with a perpendicular applied field as a function of temperature for a sample with $t_W$=0.2 nm. A diamagnetic background has been subtracted from the data at each temperature. The switching fields of the free layer (FL), synthetic antiferromagnetic layers (SAF1 and SAF2) are labeled. The inset depicts minor hysteresis loops showing only the free layer switching characteristics.

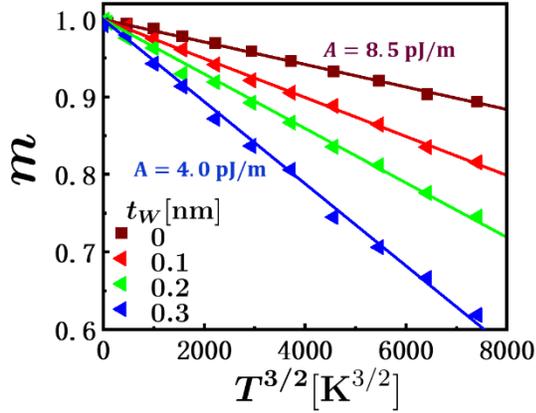

**Figure 2.** Normalized saturation magnetization $m$ versus temperature for a single CoFeB layer, MgO/CoFeB (2.3 nm)/MgO ($t_W$=0, brown symbols), and dual MgO composite CoFeB layer pMTJ stack with 0.1-0.3 nm thick W insertion layers (left pointed triangles). For pMTJ free layers, the magnetic moment is evaluated by subtracting the saturation moments of the SAF layers from the total saturation moment. Solid lines represent the fit to Bloch's law.

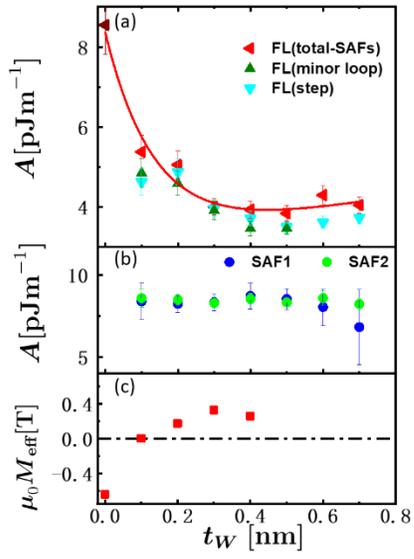

**Figure 3.** a) The exchange constant of the free layer in pMTJ layer stacks as a function of W insertion layer thickness. The solid red line is a guide to the eye. b) The exchange constant of the SAF layers, and c) the room temperature effective magnetization of the CoFeB layer from ferromagnetic resonance measurements vs. W insertion layer thickness. The error bars indicate the standard deviation from the fit. In c) the error bars are smaller than the symbols.

**Table 1.** CoFeB zero temperature magnetic moment per unit area, $M_s t$, exchange constant $A$, and effective magnetization $\mu_0 M_{eff}$, and $m(300\ K)$ for CoFeB layers. For pMTJ free layers, the magnetic moment is evaluated by subtracting the saturation moments of the SAF layers from the total saturation moment.

| $tw$ [nm] | $M_s t$ [$10^{-3}$ A] | $A$ [pJ m$^{-1}$] | $\mu_0 M_{eff}$ [T] | $m$ (300 K) |
|---|---|---|---|---|
| 0   | 2.8±0.1 | 8.5±0.9 | -0.64±0.01 | 0.92±0.01 |
| 0.1 | 2.9±0.1 | 5.4±0.4 | 0.00±0.01  | 0.88±0.01 |
| 0.2 | 2.3±0.1 | 5.1±0.5 | 0.17±0.01  | 0.82±0.01 |
| 0.3 | 1.8±0.1 | 4.0±0.3 | 0.33±0.01  | 0.72±0.01 |
| 0.4 | 2.0±0.1 | 3.9±0.3 | 0.26±0.01  | 0.74±0.01 |
| 0.5 | 1.7±0.1 | 3.8±0.4 | -          | 0.71±0.01 |
| 0.6 | 1.8±0.1 | 4.3±0.4 | -          | 0.74±0.01 |
| 0.7 | 1.6±0.1 | 4.0±0.4 | -          | 0.71±0.01 |

**Supplementary Information**

**Reduced Exchange Interactions in Magnetic Tunnel Junction Free Layers with Insertion Layers**

Jamileh Beik Mohammadi,[1] Bartek Kardasz,[2] Georg Wolf,[2] Yizhang Chen,[1] Mustafa Pinarbasi,[2] and Andrew D. Kent[1]

[1]Center for Quantum Phenomena, Department of Physics, New York University, New York, NY 10003, USA
[2]Spin Memory, Inc., Fremont, California, CA 94538, USA

**1. Magnon density and the exchange constants of the free layer**

To calculate the exchange constant using Bloch's law (Equation 1 in the main text), we need to determine the prefactor $\eta$ numerically [1]. $\eta$ is proportional to the magnon density in thermal equilibrium:

$$\eta = \frac{n_m(T)}{\left(k_B T / 2SJa^2\right)^{3/2}}, \tag{S1}$$

where $n_m(T)$ is the magnon density at temperature $T$. In cylindrical coordinates:

$$n_m(T) = \frac{1}{(2\pi)^3} \iiint \frac{k_\rho dk_\rho dk_z d\varphi}{e^{(2SJa^2(k_z^2+k_\rho^2)+hf_{FMR})/k_B T}-1}. \tag{S2}$$

Here, $k_\rho$ and $k_z$ are wave vector components in cylindrical coordinates, $S$ is the spin at each lattice site (in units of Planck's constant, $\hbar$), $J$ is the exchange energy of the ferromagnet, $k_B$ is Boltzmann constant, and $f_{FMR}$ is ferromagnetic resonance frequency at the field at which the magnetization is determined. For the free layers with strong perpendicular anisotropy, the saturation magnetization is determined from the magnetic moment near zero field (as these films have unit remanence); we therefore take the FMR frequency to be the zero applied field value. For the MgO/CoFeB/MgO sample, the applied field $H_{appl}$ needs to be larger than $H_k$ to saturate the film in the field perpendicular VSM measurements. Therefore, $f_{FMR}$ is that for an applied field greater than the saturation value.

We choose the z axis to be perpendicular to the film plane. The lateral sizes of our sample are large (2.8 mm) while the thickness of the magnetic layer $L_z$ is only a few nanometers. Therefore, for the $k_z$ component, we need to use a sum instead of an integral. Integrating Equation S1 with respect to $dk_\rho$ and $d\varphi$ replacing $\int dk_z$ with $\sum_{k_z} \Delta k_z$, we arrive at

$$n_m(T) = -\frac{1}{4\pi L_z} \frac{1}{\left(\frac{2SJa^2}{k_B T}\right)} \sum_{k_z=0}^{k_z=\pi/a} \ln\left(1 - e^{\frac{-(2SJa^2 k_z^2 + hf_{FMR})}{k_B T}}\right), \tag{S3}$$

in which, $k_z = \frac{i\pi}{L_z}$ ($i = 0,1,\ldots,\frac{L_z}{a}$), ln is the log in base $e$, and $a$ is the lattice constant of the thin film; $a$=0.287 nm for CoFeB [2]. We can determine the prefactor $\eta$ from Equation S1 by dividing the magnon density $n_m(T)$ by the argument of the Bloch law:

$$\eta = -\frac{1}{4\pi L_z} \left(\frac{k_B T}{2SJa^2}\right)^{1/2} \sum_{k_z=0}^{k_z=\frac{\pi}{a}} \left\{\ln\left(1 - e^{\frac{-(2SJa^2 k_z^2 + hf_{FMR})}{k_B T}}\right)\right\}. \tag{S4}$$

Then $\eta$ is determined numerically for the CoFeB layers. However, for the SAF1 ($t_{SAF1}$=7.6 nm) and SAF2 ($t_{SAF2}$=4.4 nm) layers, we used $\eta$ =0.080 and $\eta$ =0.110, respectively, taken from Figure S3 of reference [1]. The reason is that for our CoPt SAF layers we assume an fcc structure with $a$=0.35 similar to that of permalloy studied in reference [1].

For the free layer, we have calculated $\eta$ assuming an effective thickness of the CoFeB layer, $t_{eff} = \frac{M_s t}{1.209 \times 10^6 [Am^{-1}]}$, taking the $M_s t$ values in Table 1 in the main text. Here, $M_s$=1.209×10$^6$ Am$^{-1}$ is the zero-temperature saturation magnetization of MgO/CoFeB/MgO layer (with $t_{\text{eff}}=t_{\text{nominal}}=2.3$ nm). These values are listed in **Table S1** along with other parameters that are measured and calculated for these samples at room temperature.

## 2. Determination of the exchange constant of the free layer assuming that the magnetization is reduced throughout the CoFeB film by W

As mentioned in the main text, our VSM data shows a reduction of the zero-temperature saturation magnetization as the W insertion layer is introduced and as $t_W$ increases. In the main text, we assumed that the spin $S$ on each Co and Fe site that participates in the ferromagnetism is the same for all samples and thus that W insertion reduces the effective thickness of the CoFeB free layer, i.e. it results in a magnetic dead layer.

Another explanation for the reduced saturation magnetization is that the W layer dilutes the entire CoFeB layer leading to reduced spin on Co and Fe in the entire volume of the free layer. In this case, the CoFeB thickness is the nominal one (2.3 nm) and $S=\frac{M_s a^3}{2g\mu_B}$, where the magnetization $M_s$ is zero temperature saturation magnetization that is determined from measured saturation moment and the sample volume assuming the layer has the nominal thickness. We considered this model and calculated the saturation magnetization and the exchange constant of the free layer that are shown in **Table S2** and **Figure S2**.

This gives a lower value of $A$, and we believe this represents a lower bound on the exchange constant.

## 2. Major hysteresis loops for a pMTJ stack with a thick insertion layer

We performed magnetic hysteresis loop measurements on pMTJ stacks with dual MgO CoFeB free layers with varied W insertion layer thicknesses ($t_W$ between 0.1 and 0.7 nm). For intermediate insertion layer thicknesses ($t_W$=0.2-0.4 nm), the reversal of the free layer is a sharp step, as illustrated in Figure 1b in the main text. However, for free layers with thick insertion layer ($t_W$ larger than 0.5 nm), the exchange constant of the free layer cannot be calculated from the minor loop data and step height in major hysteresis loops. This is because for these samples, the free layer reversal occurs in two steps: one CoFeB layer switches partially and the other CoFeB layer switches gradually as the field is increased (**Figure S1**a). There are two possible reasons for such behavior. First, this can be an indication of the weak coupling between the two layers since a stronger field is needed to fully saturate the composite free layer. Secondly, the perpendicular magnetic anisotropy and $M_{eff}$ starts to decrease when $t_W$ is greater than 0.3 nm (as shown in Figure 3b in the main text). As a result, for thicker insertion layers, higher field would be required to saturate the free layer in the field perpendicular geometry. Therefore, $M_s$ and exchange constant of the free layer cannot be evaluated from the minor loop hysteresis curves. In this case, the magnetic moment of the pMTJ free layer has to be evaluated by subtracting the saturation moments of the SAF layers from the total saturation moment. This explains the uncertainties in determining the exchange constant of the free layers with thick insertion layer (Figure 3a in the main text and Figure S2 in this supplementary information).

In addition, the exchange constant of the free layer that is calculated from the different methods does not agree for the thinnest insertion layer ($t_W$=0.1 nm). One reason for this disagreement is that for this sample, the effective perpendicular anisotropy (as shown in Figure 3c) is not sufficient to produce an out-of-plane easy axis. Therefore, the reversal of the free layer is not a sharp single step, and higher perpendicular fields are required to fully saturate the layer (Figure S1b). As a result, $M_s$ of this free layer cannot be estimated accurately from the minor loop or step height methods that are explained in the main text and the method of subtracting the SAF moments from the full hysteresis loop is used in this case.

For pMTJ stacks with $t_W$=0.2-0.4 nm, the magnetization reversal of individual layers is well-spaced in field. This allows us to extract exchange constant of all the ferromagnetic layers including the SAF layers. As a result, the free layer exchange constant that is calculated from three different methods (that are explained in the main text) agree well.

**4. Calculation of exchange constant for NiFe sample**

We performed magnetic measurements on a 120 nm NiFe film provided by the National Institute of Standard and Technology (NIST) to check our methodology, one of the series of samples studied in Reference [1]. We used Bloch's law to fit this data and evaluate the exchange constant of this film. Saturation magnetization vs. $T^{3/2}$ data and the fit to Bloch's law is plotted in **Figure S3**. Using η=0.058 from Figure S3 of Reference [1], we calculated the exchange constant $A$(300 K)=7.8 pJm$^{-1}$ which is consistent with the data presented in Figure 4c of this reference.

**References**

[1]     H. T. Nembach, J. M. Shaw, M. Weiler, E. Jué, T. J. Silva, *Nat. Phys.* **2015**, 11, 825.

[2]     A.K. Kaveev, V.E. Bursian, B.B. Krichevtsov, K.V. Mashkov, S.M. Suturin, M.P. Volkov, M. Tabuchi, and N.S. Sokolov, *Phys. Rev. Mater.* **2018**, 2, 014411.

[3]     S. Miura, T.V.A. Nguyen, Y. Endo, H. Sato, S. Ikeda, K. Nishioka, H. Honjo, and T. Endoh, *IEEE Trans. Magn.* **2019**, 1,2901841.

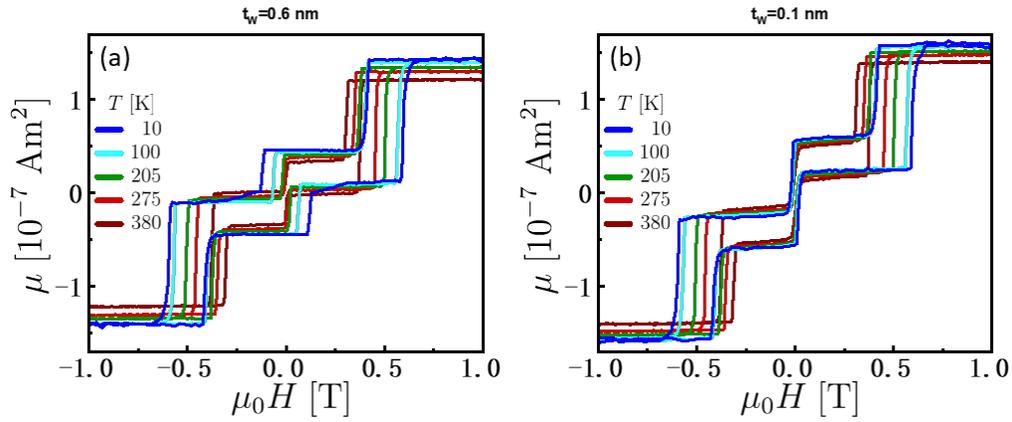

**Figure S1.** Major hysteresis loops for pMTJ stack with a) $t_W$=0.6 nm, and b) $t_W$=0.1 nm insertion layer in temperature range of 5-380 K.

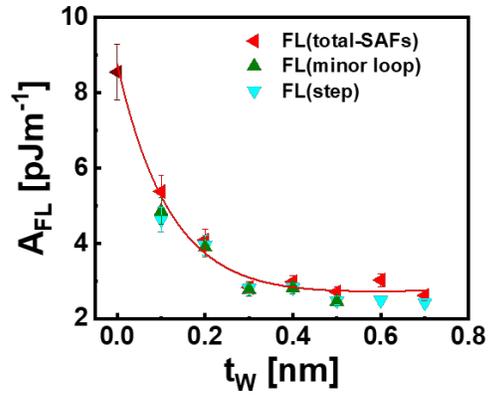

**Figure S2.** Exchange constant of the CoFeB layer at zero temperature from three methods that are explained in the main text. For the data presented in this figure, we have used $t_{nominal}$=2.3 nm and $\eta$ values that are listed in Table S2. Solid line is guide to eye.

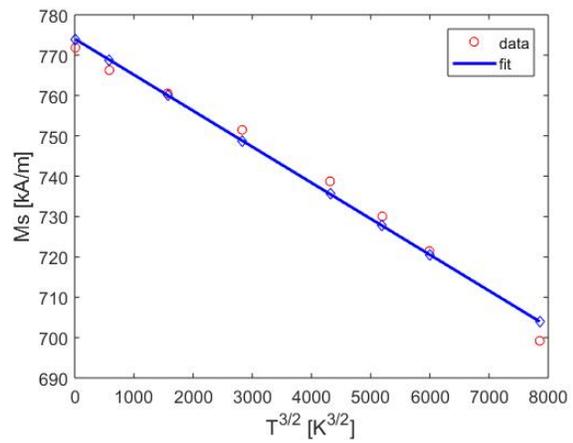

**Figure S3**. $M_s$ vs. $T^{3/2}$ data (red symbols) and fit to the Bloch's law (blue symbol and blue solid line) for 120 nm NiFe.

**Table S1.** CoFeB layer nominal thickness, effective thickness, insertion layer thickness, room temperature saturation magnetization, effective magnetization, room temperature exchange constant (the zero temperature exchange constant is listed in Table 1 in the main text), and the prefactor η considering $t_{eff}$.

| $t_{nominal}$ [nm] | $t_{eff}$ [nm] | $t_W$ [nm] | $\mu_0 M_{eff}$ [T] | $A(300\ K)$ [pJm$^{-1}$] | η (based on $t_{eff}$) |
|---|---|---|---|---|---|
| 2.3 | 2.3 | 0   | -0.64±0.01 | 7.3±0.7 | 0.083 |
| 2.3 | 2.4 | 0.1 | 0.00±0.01  | 4.2±0.4 | 0.074 |
| 2.3 | 1.9 | 0.2 | 0.17±0.01  | 3.1±0.2 | 0.089 |
| 2.3 | 1.5 | 0.3 | 0.33±0.01  | 1.9±0.1 | 0.095 |
| 2.3 | 1.6 | 0.4 | 0.26±0.01  | 1.9±0.1 | 0.089 |
| 2.3 | 1.4 | 0.5 | -          | 1.8±0.1 | 0.095 |
| 2.3 | 1.5 | 0.6 | -          | 2.1±0.1 | 0.094 |
| 2.3 | 1.3 | 0.7 | -          | 1.8±0.1 | 0.099 |

**Table S2.** CoFeB layer saturation magnetization $M_s$, exchange constant $A(300\ K)$, and effective magnetization $\mu_0 M_{eff}$, g-factor, and the pre-factor η. The values listed in this table are considering the nominal thickness of $t_{nominal}$=2.3 nm. For pMTJ free layers, the magnetic moment is evaluated from subtracting the saturation moments of the SAF layers from the total saturation moment.

| $t_{CoFeB}$ [nm] | $t_W$ [nm] | $M_s(0\ K)$ [kAm$^{-1}$] | S | $M_s(300\ K)$ [kAm$^{-1}$] | $\mu_0 M_{eff}$ [T] | g-factor | $A$ [pJm$^{-1}$] | $A(300\ K)$ [pJm$^{-1}$] | η |
|---|---|---|---|---|---|---|---|---|---|
| 2.3 | 0   | 1209±3 | 0.73 | 1118±20 | -0.64+0.01 | 2.10±0.01 | 8.5±0.8 | 7.3±0.7 | 0.083 |
| 2.3 | 0.1 | 1260±5 | 0.74 | 930±11  | 0.00±0.01  | 2.15±0.01 | 5.4±0.5 | 4.2±0.4 | 0.074 |
| 2.3 | 0.2 | 990±9  | 0.58 | 793±8   | 0.17±0.01  | 2.12±0.02 | 4.1±0.3 | 2.7±0.3 | 0.073 |
| 2.3 | 0.3 | 784±7  | 0.46 | 566±6   | 0.33±0.01  | 2.17±0.07 | 2.8±0.3 | 1.5±0.2 | 0.071 |
| 2.3 | 0.4 | 865±15 | 0.48 | 604±6   | 0.26±0.01  | 2.26±0.03 | 3.0±0.3 | 1.6±0.2 | 0.071 |
| 2.3 | 0.5 | 763±9  | 0.43 | 490±5   | -          | -         | 2.7±0.2 | 1.4±0.2 | 0.072 |
| 2.3 | 0.6 | 795±7  | 0.44 | 528±6   | -          | -         | 3.0±0.2 | 1.7±0.2 | 0.072 |
| 2.3 | 0.7 | 684±7  | 0.38 | 434±5   | -          | -         | 2.6±0.2 | 1.3±0.2 | 0.073 |